\documentstyle[epsfig]{mn}
\title[Spectral lags in gamma-ray bursts]{Spectral lags and the energy dependence
of pulse width in gamma-ray bursts: contributions
from the relativistic curvature effect}
\author[R.-F. Shen, L.-M. Song \& Z. Li]
{Rong-feng Shen
\thanks{E-mail: rfshen@astro.as.utexas.edu. Present address:
Department of Astronomy, University of Texas at Austin, Austin, TX 78712, USA},
Li-ming Song, and Zhuo Li\\
Particle Astrophysics Center, Institute of High Energy Physics,
Chinese Academy of Sciences, Beijing 100039, China}

\date{Accepted 2005 April 27. Submitted 2004 July 20}
\begin{document}

\maketitle

\begin{abstract}
We compute the temporal profiles of the gamma-ray burst pulse in the four
Burst and Transient Source Experiment (BATSE) Large Area Detector (LAD)
discriminator energy channels, with the relativistic curvature
effect of a expanding fireball being explicitly investigated.
Assuming an intrinsic ``Band'' shape spectrum and an intrinsic
energy-independent emission profile, we show that merely the
curvature effect can produce detectable spectral lags if the
intrinsic pulse profile has a gradually decaying phase. We examine the
spectral lag's dependences on some physical parameters, such as
the Lorentz factor $\Gamma$, the low-energy spectral index, $\alpha$, of
the intrinsic spectrum, the duration of the intrinsic radiation
$t_d'$ and the fireball radius $R$. It is shown that
approximately the lag $\propto \Gamma ^{-1}$ and $\propto t_d'$,
and a spectrum with a more extruded shape (a larger $\alpha$)
causes a larger lag. We find no dependence of the lag on $R$.
Quantitatively, the lags produced from the curvature effect are
marginally close to the observed ones, while larger lags require
extreme physical parameter values, e.g., $\Gamma < 50$, or $\alpha
> -0.5$. The curvature effect causes an energy-dependent pulse
width distribution but the energy dependence of the pulse width we
obtained is much weaker than the observed $W \propto E^{-0.4}$
one. This indicates that some intrinsic mechanism(s), other than the
curvature effect, dominates the pulse narrowing of GRBs.
\end{abstract}


%%%%%%%%%%%%%%%%%%%%%%%%%%%%%%%%%%%%%%%%%%%%%%%%%%%%%%%%%%%%%%%%%%%%%%%%%%%%%%
\section{Introduction}

Cheng et al. (1995) were the first to analyse the spectral lag of
gamma-ray bursts (GRBs), which is the time delay between the peaks
in the Burst and Transient Source Experiment (BATSE) Large Area
Detector (LAD) Channel 1 (25 - 50 keV) and Channel 3(100 - 300
keV) light curves. Subsequently, several authors have carried out
more analysis work on the GRB lags. Norris et al. (1996) and
Norris, Marani \& Bonnell (2000) found the cross-correlation
function lags between BATSE Channel 1 and Channel 3 photons tend
to concentrate near $< 100$ ms; for six bursts with known
redshift, the $z$-corrected lags are distributed between 6 and 200
ms. Wu \& Fenimore (2000) extended the analysis to very low energy
($\sim$ 2 keV); they found that about $20\%$ of GRBs have
detectable lags and that GRBs don't show larger lags at lower
energy. A recent measurement by Chen et al. (2005) for the BATSE
bursts shows that the majority of the lags are below $\sim$ 200
ms, and that the histogram of the lags peaks around 30 ms. More
intriguingly, Norris et al. (2000) found that, for those six
bursts with known $z$, the peak luminosity is anti-correlated with
the lags. This relationship provides a useful tool to estimate the
distances of large sample of GRBs by analysing their light curves.

Three theoretical explanations for the lag/luminosity relation have been proposed:
the relationship is due to the variation in line-of-sight velocity among bursts
(Salmonson 2000); it is caused by the variation of the off-axis angle when viewing a
narrow jet (Ioka \& Nakamura 2001); or it is caused by radiation cooling - highly
luminous burst
cools fast and the lag will be short (Schaefer 2004). However, the problem of
what mechanism(s) causes the spectral lags of GRBs remains unresolved. Salmonson (2000)
did not explain the origin of the lag but assumed that it derives from some proper decay
time scale $\Delta t'$ in the rest frame of the emitter. In the model by
Ioka \& Nakamura (2001),
the lag is caused by the far side of the emitting region producing lower-energy radiation
after a longer light-travel time, for a narrow jet with viewing angles outside the cone
of jet. However, their model requirements seem too stringent (see below). Schaefer (2004)
proposed radiative cooling as the origin of the lag. The difficulty of this explanation
is that in order to adjust the observed synchrotron cooling time scale to be comparable
with the lag time scale, the magnetic field has to be $\sim 7$ Gauss, a value much below
the strength required
by most of the current models (e.g. Piran 1999).

Kocevski \& Liang (2003) have assumed that the observed lag is the direct result of
spectral evolution, another property of GRBs (Norris et al. 1986; Bhat et al. 1994).
In particular, as the peak energy of the GRB's $\nu F_{\nu}$ spectrum decays through the
four BATSE channels, the photon-flux peak in each individual channel will
be shifted, probably producing the measured lag.
From a sample of 19 GRBs, Kocevski \& Liang (2003) found an empirical relation
between the peak energy's decaying rate and the GRB lag.

It is widely accepted that the gamma-rays come from a relativistically expanding
fireball surface with Lorentz factor $\Gamma >$ 100 (Lithwick \& Sari (2001) and references
therein). At some distance from the central source (e.g. $R= 10^{12}\sim10^{14}$cm, {\it cf.}
Piran (1999)), photons emitted from the region on the line
of sight and those from the side region at an angle of $\theta \sim 1/\Gamma$
with respect to the line of sight are Doppler-boosted by different factors and travel
different distances to the observer. This is what we call the curvature effect.
Comparing the radiation from the side region and that from the line-of-sight region,
for the latter its photons are  Doppler-boosted to higher energies and arrive to the
observer earlier; its observed temporal structure will be boosted to be narrower.

The motivation of this paper is to see how the curvature effect will change the
intrinsic pulse profile in different energy channels, and special interest is
focused on whether merely the curvature effect can produce the spectral lags of the pulses.
Except for the soft photon lags, the pulses in GRBs show another temporal
property, i.e. the pulse narrowing at higher energy, or pulse width as a function
of energy (Fenimore et al. 1995, Norris et al. 1996). We are also interested in probing the
contributions of the curvature effect to these properties.

A fireball internal-external shocks model has emerged for the theoretical understanding
of the origin of GRBs (Piran 1999, 2004). According to this model, GRBs are produced when
an ultra-relativistic outflow dissipates its kinetic energy through the internal collisions
within the outflow itself. The afterglow occurs when the flow is decelerated by shocks
with the circumburst medium. This model has made many successful explanations and predictions
to the observations of GRBs. Our investigation to the curvature effect will be based on
the frame of the internal shock model, where the shock is generated from the colliding
shells, and electrons are accelerated in the shock and radiate.

Ioka \& Nakamura (2001) proposed a model in which a narrow jet is viewed at off-axis
angle to
explain the lag/luminosity relation and the variability/luminosity relation. Their
model successfully reproduces the lag/luminosity relation, while the lag is caused
by the curvature effect of the jet, which increases with the off-axis angle. However, one
substantial problem with this model, as pointed out by Schaefer (2004), is that
it works with an exacting assumption that the jet opening angle always equals
$\Gamma^{-1}$. Furthermore, Ioka \& Nakamura (2001) only consider an instantaneous
emission in
the jet rest frame, which is a too simple assumption. Consequently, in their model there
is no lag when the jet is viewed near-on-axis.
Different from Ioka \& Nakamura (2001), we have considered various rest-frame emission profiles
and assumed an isotropically expanding radiation surface.

We present our model in Section 2, including the basic assumptions and formulas. The major
results are presented in Section 3, based on which we give our conclusions and discussion
in Section 4.

%%%%%%%%%%%%%%%%%%%%%%%%%%%%%%%%%%%%%%%%%%%%%%%%%%%%%%%%%%
\section{MODEL}

\subsection{Three time scales}

Three time scales are involved in determining the temporal structures of pulses
in GRBs: (i)cooling time scale; (ii)hydrodynamic time scale; (iii)angular spreading
time scale (Kobayashi et al. 1997, Wu \& Fenimore 2000).

In the synchrotron cooling model, the shock-accelerated electrons
cool via synchrotron emission,
and the electron's average energy becomes smaller and the radiated power decays.
As pointed out by Wu \& Fenimore (2000) , the standard internal-shock model
gives an observed synchrotron cooling time scale at a given photon energy as
$$T_{syn}(h\nu) \sim 2\times 10^{-6} s \,\,
\epsilon_B^{-3/4} (\frac{h\nu_{obs}}{100 keV})^{-1/2},$$
where $\epsilon_B$ is the equipartition parameter for the ratio of the magnetic energy
density to the total internal energy density; for a typical value in the internal-shock
model, e.g. $\epsilon_B=0.01$ ($B\sim 10^5$ G), the cooling time scale is
far shorter than the lag time scale.

The hydrodynamic time scale is related to
the energizing of the electrons. In the internal-shock model,
if one assume the the local microscopic acceleration of electrons is instantaneous,
then the hydrodynamic time scale is attributed to the shell-crossing time of the shock,
$T_{dyn}'=\Delta'/v_{sh}'$, where $\Delta'$ is the shell width and $v_{sh}'$ is
the shock velocity, both in the comoving frame of the upstream flow.
We hardly know about $\Delta'$. However if one assume the shells are radially
expanding, in the observer frame this time scale is
$$T_{dyn} \sim 1 s \,\, \beta_{sh}'^{-1}
(\frac{R}{10^{15}cm})(\frac{\Gamma}{100})^{-2},$$
where $R$ is the radius at which the shell radiates, and $\Gamma$ is the Lorentz
factor of the shell (Ryde \& Petrosian 2002).
(Note that apart from the shock acceleration scenario,
there can be other particle energizing mechanisms, e.g. magnetic field
reconnection (Stern 1999), which may have a different hydrodynamic time scale)

The angular spreading time scale is the delay between the
 arrival times of the photons emitted at the line-of-sight region and of
that emitted at the side region of the shell (Sari \& Piran 1997).
Because of the relativistic
beaming of the moving radiating particles, only the emission from a narrow
cone with an opening angle of $\sim 1/\Gamma$ is observed. This gives a time
scale of the delay of
$$T_{ang}=1.7 s \,\, (\frac{R}{10^{15}cm})(\frac{\Gamma}{100})^{-2}.$$

\subsection{Assumptions}

In this paper, different from Schaefer (2004), we assume the cooling time scale
is much shorter than the other two time scales, i.e. the accelerated particles
radiate their energy rapidly. Thus the rest frame duration of the emission is
determined by the hydrodynamic time scale.

We consider a thin shell expanding with a relativistic speed, whose Lorentz factor is
$\Gamma$. The shell begins to radiate at radius $R$.
In the co-moving frame of the shell, the radiation intensity of the shell surface
$I'(\nu',t')$ is assumed to be isotropic, and has an energy-independent time history
$f'(t')$. Note that all the quantities in the rest frame of the radiation
surface are labeled with a prime note.

Band et al. (1993) found that the GRB spectra are well described at low
energy by a power law with an exponential cutoff and by a steeper power
law at high energy. The typical fitted value distributions for the low-energy spectral
index ($\alpha$) is -1.5 $\sim$ -0.3, the high-energy spectral index ($\beta$)
is -2 $\sim$ -3
and the peak energy ($E_p$) of the $\nu F_{\nu}$ spectrum is 100 $\sim$ 500 keV
(see Preece et al. 2000).
By modeling, Qin (2002) showed that the relativistic expanding of the fireball
does not alter the shape of GRB's rest frame spectrum, but only shifts the peak of
the spectrum to a higher energy; the distribution of $E_p$ is scaled with the
Lorentz factor $\Gamma$, as shown in their Table 4. So we directly adopt the spectral
form used by
Band et al. (1993) as the rest frame spectrum, and choose values for the low-energy
spectral index $\alpha$ to be -0.8 and the high-energy spectral index $\beta$ be -2.4;
The peak
energy of the rest frame spectrum $E_p'$ is adjusted such that $E_p$ is 350 keV,
where the relation $E_p = 1.65\times \Gamma E_p'$, derived from Table 4 in
Qin (2002) in the range of $10<\Gamma<2000$, is used.

\subsection{The formulas}

The radiation intensity in the observer's frame is $I(\mu,\nu,t) = I(\mu,\nu)f(t)$. It is
connected with the rest frame radiation intensity by
\begin{equation}
I(\mu,\nu)f(t) = {\cal D}^3(\mu)I'(\nu')f'(\Gamma t')
\end{equation}
and
\begin{eqnarray} \nu = \nu'{\cal D}(\mu) , \end{eqnarray}
where $\mu = {\rm cos}\theta$, and $\theta$ is the angle of the concerned local radiation
surface to the line of sight; ${\cal D}(\mu)=[\Gamma(1-\mu\beta)]^{-1}$ is the local
relativistic Doppler factor respect to the observer, where $\Gamma$ is the Lorentz factor
of the radiation surface and $\beta = \sqrt{1-\Gamma^{-2}}$.

We use $t$ to refer to the photon emitting time and use $T$ to refer to the time
that the photon arrives at the observer. We define that, at the time $t=0$, the first
photons are emitted from the surface and, for simplicity, we also tune the arrival
time $T$ of the first photon emitted from the line-of-sight region ($\mu =1$) to be 0 too.
Then the arrival time of a photon emitted from region $\mu$ at
time $t$ is
$$T = (1-\beta)t + (1-\mu)(R+\beta ct)/c.$$
The first part of the right-hand side of the equation is caused by the motion of the shell,
and the second part is due to the difference between the light travel distances of photons
emitted from the line-of-sight region and from the side region.
It can be rewritten as
\begin{eqnarray}
T = (1-\mu\beta)t+(1-\mu)\tau , \label{relat}
\end{eqnarray}
where $\tau=R/c$, which connects the photon's arrival time, emitting time and emitting
place in one equation.

At time $T$, the observed flux comes from the photons emitted at the region
$1>\mu>\mu(T,t=0)$, where the boundary $\mu(T,t=0)=1-T/\tau$, calculated from
equation (\ref{relat}), is the place whose first photon is emitted with arrival
time $T$. The observed specific flux can be obtained by integrating the radiation
intensity over this region
\begin{eqnarray}
  F_{\nu}(T) &=&  -\int_{1}^{1-T/\tau}I_{\nu}(\mu,t)\mu R^2(t) d\mu  \label{sf}
\end{eqnarray}

For a BATSE LAD discriminator channel ($\nu_a, \nu_b$), the observed photon counts flux is
$$n_{ab}(T) = \int_{\nu_a}^{\nu_b} \frac{F_\nu(T)}{\nu} d\nu .$$
From equation (\ref{sf}), substituting the emitting time $t$ in the integral with $T$
and $\mu$ using equation (\ref{relat}), $n_{ab}(T)$ can be rewritten as
\begin{eqnarray*}
n_{ab}(T)& = & -[\tau+(T-\tau)\beta]^2 \int_{1}^{1-T/\tau} \\ & &
\times \frac{f[t(T,\mu)]}{(1-\mu\beta)^2} \mu d\mu {\cal D}^3(\mu)
\int_{\nu_a / {\cal D}(\mu)}^{\nu_b / {\cal D}(\mu)} \frac{I'_{\nu'}}{\nu'} d\nu' ,
\end{eqnarray*}
where $R(t)=c[\tau+t(T,\mu)\beta]$ and equations (1), (2) are used.
So we get the explicit formula that is used in our calculation.

\subsection{Intrinsic time profile of the radiation}

Here we introduce three cases of the intrinsic emission time profile.

(i) Rectangular profile. During a finite duration, the emissivity is constant,
with the instantaneous rising phase and decaying phase:

\begin{displaymath}
f(t)= \left\{\begin{array}{ll} 1, & 0<t<t_d \\ 0, & t>t_d
\end{array}\right.
\end{displaymath}
In our assumptions $t_d$ is
related to the hydrodynamic time scale discussed above, but note
that $t_d$ is defined in the observer's frame, without taking into
account the Doppler boosting (i.e., it is not the observed time
scale). In addition, we refer to $t_d'$ as the $t_d$ measured in the
comoving frame of the shell.

(ii) One-sided exponential profile. The emissivity decays exponentially after
an instantaneous rising phase:

\begin{displaymath}
f(t) = \left\{\begin{array}{ll}
       0, & t<0 \\ \exp[-(t/t_d)], & t>0 \end{array}\right.
\end{displaymath}

(iii) Symmetric Gaussian profile. The emissivity has both a finite-time rising
phase and a finite-time decaying phase:

\begin{displaymath}
 f(t) = \left\{\begin{array}{ll}
0, & t<0 \\ \exp[-(\frac{t-1.5t_d}{t_d})^2], & t>0
\end{array}\right.
\end{displaymath}
where we introduce the coefficient -1.5 in order that the emission starts at $t=0$
with the radiation intensity of a tenth ($e^{-1.5}\approx 0.1$) of its peak value.

%%%%%%%%%%%%%%%%%%%%%%%%%%%%%%%%%%%%%%%%%%%%%%%%%%%%%%%%%%
\section{RESULTS}

\subsection{Observed temporal profiles in different energy bands and time lags}

\begin{figure}
\centerline{\hbox{\psfig{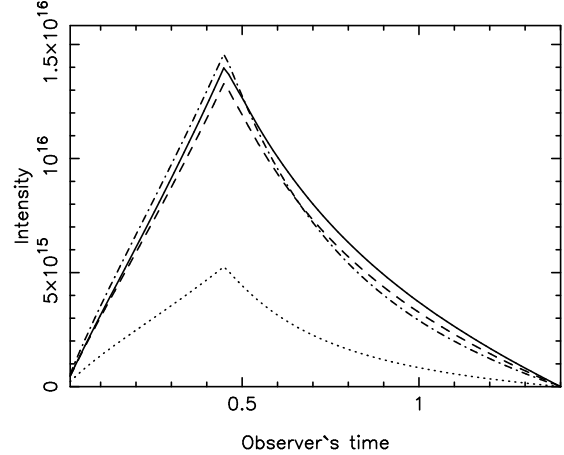}}}
\caption {Observed pulse from intrinsic rectangular time profile
in four BATSE LAD energy channels. {\em Solid line}: 25 -- 50 keV; {\em
dashed line}: 50 -- 100 keV; {\em dot-dashed line}: 100 -- 300 keV; {\em
dotted line}: 300 -- 1800 keV. The vertical coordinate is in
arbitrary units. The peaks of the pulses at 4 energies arrived
simultaneously. $\Gamma$ = 100, $\alpha$ = -0.8, $\beta$ =- 2.4,
$R=3\times10^{14}$ cm, $t_d=0.9\times10^4$ s. The rising phase comes
from the expanding of the radiation surface and the rising time is
determined by duration of the intrinsic radiation. The decay phase
is due to the angular spreading effect.} \label{recfig}
\end{figure}

\begin{figure}
\centerline{\hbox{\psfig{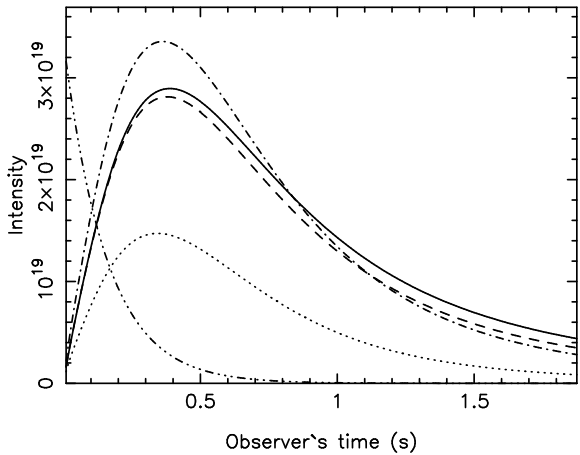}}}
\caption{Observed pulse from one-sided exponential intrinsic time
profile in four BATSE LAD energy channels. {\em Solid line}:
25 -- 50 keV; {\em dashed line}: 50 -- 100keV; {\em dot-dashed line}:
100 -- 300 keV; {\em dotted line}: 300 -- 1800 keV; {\em
triple-dots-dashed line}: the intrinsic emission profile. The
vertical coordinate is in arbitrary units. $\Gamma$ = 400,
$\alpha$ = -0.8, $\beta$ = -2.4, $R=5\times10^{14}$ cm,
$t_d=5\times10^4$ s. Lag$_{13} \simeq$ 0.03 s.}  \label{expfig}
\end{figure}

\begin{figure}
\centerline{\hbox{\psfig{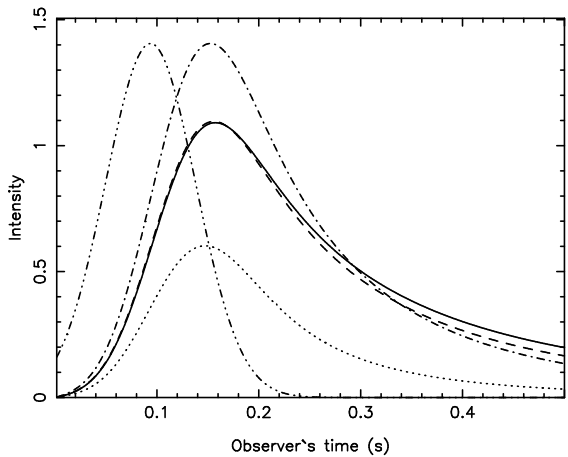}}}
\caption{Observed pulse from symmetric Gaussian intrinsic time
profile in four BATSE LAD energy channels. {\em Solid line}:
25 -- 50 keV; {\em dashed line}: 50 -- 100 keV; {\em dot-dashed line}:
100 -- 300 keV; {\em dotted line}: 300 -- 1800 keV; {\em
triple-dots-dashed line}: the intrinsic emission profile. The
vertical coordinate is in arbitrary units. $\Gamma$ = 400,
$\alpha$ = -0.8, $\beta$ = -2.4, $R=5\times10^{14}$ cm,
$t_d=2\times10^4$ s. Lag$_{13}\simeq$ 0.005 s.}  \label{gaufig}
\end{figure}

Using the equations and the typical parameters introduced above,
we calculate the observed pulse profiles in the four BATSE LAD energy
channels, assuming three different intrinsic pulse profiles.
The results are plotted in Figure \ref{recfig},
\ref{expfig} and \ref{gaufig}, respectively.

First, we calculate the pulse light curves in the four energy channels for the
rectangular intrinsic radiation profile. It has a steady rising phase,
followed by a distinct peak, as shown in Figure \ref{recfig}. The rising of
the flux is due to the expanding of the radiation surface. The rising is steady
because the radiation intensity is constant during a finite time. The decay phase
of the observed pulse is due to the angular spreading effect. The peak occurs
when the intrinsic radiation begins to cease. For various parameter spaces
(duration of the emission $t_d: 10^3 - 10^5$ s, radius $R: 10^{13}-10^{15}$ cm,
Lorentz factor $\Gamma: 50-500$),
we do not detect any time lag between the peaks observed at different
energies in this case.

As for the one-sided exponential and the symmetric Gaussian
emission profiles, their observed light curves calculated in the four
energy channels are illustrated in Figure \ref{expfig} and
\ref{gaufig}, respectively. For these two emission profiles, their
peaks are more gradual. The maximum of the photon flux at higher
energy arrives early than at lower energy. If we define the time
lag as being the difference between the arrival times of the peaks
in energy Channel 1 and 3 or 4, the reproduced lags are
quantitatively comparable to the observed ones (e.g., $\sim
10^{-2}$ s if corrected for the cosmological time dilation. cf.
Norris et al. (2000)). Note that here we do not take into account
the redshifts, $z$, of the GRBs, which must make the observed lags
$(1+z)$ times larger.

The intrinsic rectangular profile cannot produce the observed lags
because in this case the radiation diminishes immediately. In the
observed pulse, the rising phase comes from the expanding of the
radiation surface and the rising time is determined by duration of
the intrinsic radiation ($\approx t_d / (2\Gamma^2)$; also see
Qin et al. 2004). The decay phase is due to the angular
spreading effect. Hence if the intrinsic radiation switches off
immediately, the transition from the rising phase to the decay
occurs abruptly and induces a sharp peak in the observed pulse.
For this case, the immediate switch-off of the intrinsic radiation
dilutes the relativistic curvature effect, and hence does not
produce the peak lags.

\subsection{The pulse width - energy relation}

For the three intrinsic emission profiles, we find that the pulse observed in
high-energy channel is narrower than in lower-energy channel, which is manifested
by the fact that the FWHMs of the pulses in separated energy channels decrease
with the energy in a power-law form. However, the pulse narrowing we obtained is
less prominent than that observed in real GRBs for which pulse width decay
power-law index $\sim$-0.4; while the pulse width decay index we obtained,
for instance in the case of Figure \ref{expfig} (one-sided exponential decay
emission profile), is -0.13.

Other than the ``Band'' spectrum, we also used an alternative function - a low-energy
power law plus the high-energy exponential cut-off at $E_p'$ - for the rest-frame
spectrum. For the typical values of the parameters we have used, this changing of
spectrum only narrows the Channel 4 pulse by $\sim$ 6\%, hence hardly changes the
slope of the pulse width versus the energy.

\subsection{Lag's dependence on other physical parameters}

The spectral lag is an important observational property of the pulse in GRBs in that it
may be used to derive the cosmological distribution of GRBs (Norris 2002) and to
discriminate the internal shock signature and the external shock signature in the
pulses (e.g., Hakkila \& Giblin 2004). This motivates us to probe the dependences of the
peak lags on other physical parameters of the simple model. We choose the symmetric
Gaussian profile as the intrinsic emission profile, which includes an intrinsic
rising phase. We alter the Lorentz factor $\Gamma$, the spectral parameters of the rest
frame emission ($\alpha$, $\beta$ and $E_p'$), the radius of the radiation surface $R$,
and the rest-frame duration $t_d'$ of the intrinsic radiation, respectively, and see
how the Channel 1/3 and the Channel 1/4 peak lags vary with these changes.

\subsubsection{Lorentz factor}

\begin{figure}
\centerline{\hbox{\psfig{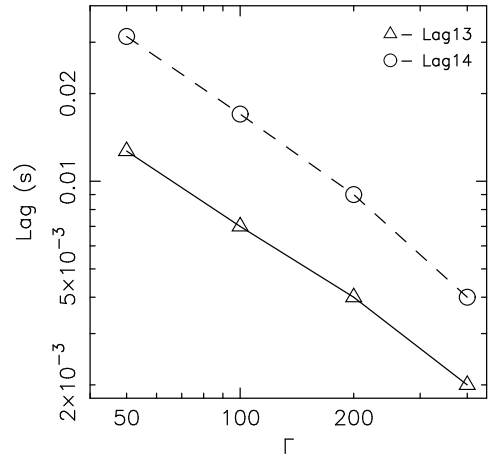}}}
\caption{Symmetric Gaussian intrinsic pulse: lag dependence on
the Lorentz factor. $\alpha$ = -0.8, $\beta$ = -2.4,
$R=5\times10^{13}$ cm, $t_d^{'}$ = 20 s. Power-law fit gives index
-0.87 and -0.93, respectively.} \label{lag_gamma}
\end{figure}

Figure \ref{lag_gamma} shows that the lag decreases with the Lorentz factor following
Lag $\propto \Gamma^{-1}$. We think this is a natural outcome of the relativistic boosting
of the time structure. Those pulses whose lags are larger may come from colliding shells
with low Lorentz factors, according to the current standard models (e.g., Piran 1999).

\begin{figure}
\centerline{\hbox{\psfig{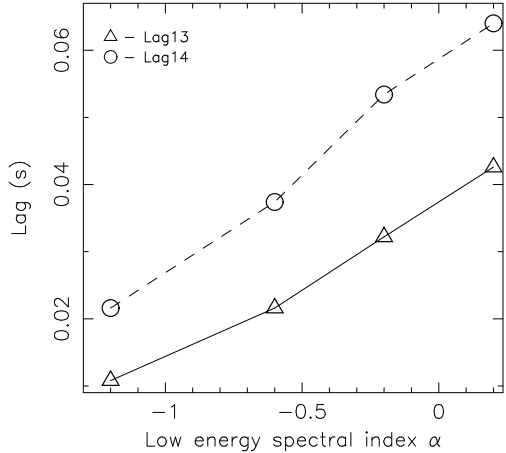}}}
\caption{Symmetric Gaussian intrinsic pulse: lag dependence on
$\alpha$, the low-energy spectral index of the adopted spectrum.
$\Gamma$ = 100, $E_p^{'}$= 1.75 keV, $\beta$ = -2.4,
$R=5\times10^{13}$ cm, $t_d^{'}$ = 40 s. Note that GRB spectrum with
$\alpha > 0$ is very rarely observed (Preece et al. 2000); the
calculated data points there are only used to show the
tendency.} \label{lag_alpha}
\end{figure}

\begin{figure}
\centerline{\hbox{\psfig{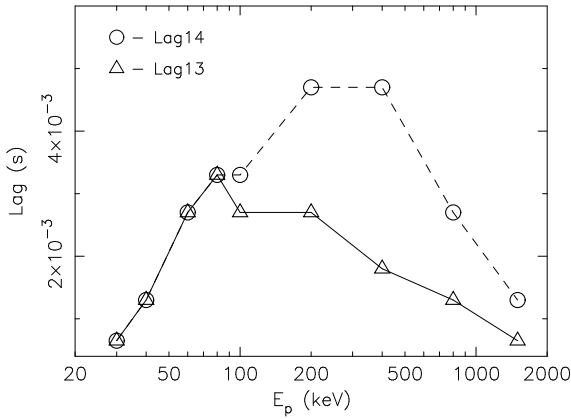}}}
\caption{Symmetric Gaussian intrinsic pulse: lag dependence on
the observed spectral break energy $E_p$ of the pulse spectrum.
$\Gamma$= 400, $\alpha$= -0.8, $\beta$= -2.4, $R$= 5$\times10^4$ cm,
$t_d'$= 100 s. Note that $E_p$ is obtained through a simple scaling
relation, $E_p = 1.65\times \Gamma E_p'$, between $E_p$ and the
break energy, $E_p'$, of the rest-frame emission spectrum, as a
result of the simulation by Qin (2002).} \label{lag_ep}
\end{figure}

\subsubsection{Spectral parameters}

Figure \ref{lag_alpha} shows that the lag increases with the low energy spectral index
$\alpha$ of the rest frame spectrum of the pulse. A larger $\alpha$ means more photons
are concentrated around the peak energy $E_p'$ of the $\nu'F_{\nu}'$ spectrum. Then,
for a narrower spectrum, the curvature effect will work more effectively in producing
the spectral lags. This conjecture is supported when we alter the high-energy
spectral index $\beta$. We found that a steeper high-energy power-law spectrum produces
a larger lag, e.g., the Channel 1/4 lag has a 14\% increase for $\beta$
changing from -2.4 to -3.0. Compared with the Channel 1/4 lag, the Channel
1/3 lag has a weaker dependence upon $\beta$.

The lag's dependence on the observed break energy of the spectrum $E_p$ is
shown in Figure \ref{lag_ep}. The lag has its maximum when $E_p$ falls near the starting
energy of the corresponding high-energy channel that is used in measuring the lag
(i.e., $\sim$ 100 keV for Channel 3, $\sim$ 300 keV for Channel 4).

The above findings about the lag's dependences on the spectral parameters are
qualitatively consistent with the tendency observed in those long-lag wider-pulse
bursts by Norris et al. (2005). They found that their long-lag
(measured for Channel 1/3) burst sample has, on average,
lower $E_p$ (centered around $\sim 110$ KeV), larger $\alpha$
({\it harder} low-energy power law) and smaller $\beta$
({\it softer} high-energy power law),
than the bright burst sample analyzed by Preece et al. (2000).

Substituting the ``Band'' spectrum with an alternative one of a single power law plus
an exponential high-energy cut-off causes no changes to the Channel 1/3 lag, while
the Channel 1/4 lag has a $\sim$40\% increase if the observed cut-off energy $E_p$
is below 300 keV; for $E_p >$ 300 keV, the increase of Channel 1/4 lag is much smaller.

\subsubsection{Duration of the emission}

\begin{figure}
\centerline{\hbox{\psfig{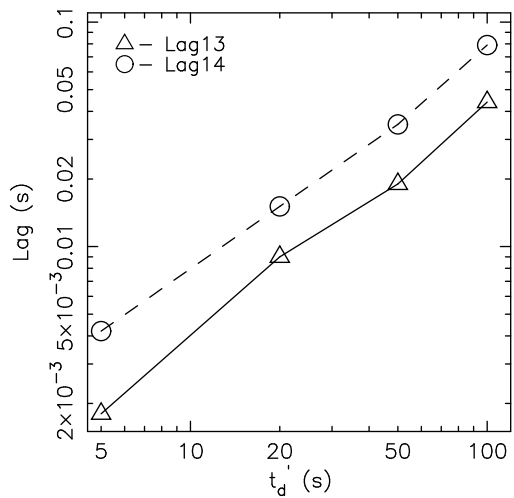}}}
\caption{Symmetric Gaussian intrinsic pulse: lag dependence on
$t_d'$, the rest frame duration of the radiation. $\Gamma$= 100,
$E_p^{'}$= 1.75 keV, $\alpha$= -0.8, $\beta$= -2.4,
$R$=5$\times10^{13}$ cm.}  \label{lag_td'}
\end{figure}

We find that longer rest-frame duration ($t_d'$) of emission will cause larger lags,
as is shown in Figure \ref{lag_td'}. This result may be associated with an observed
tendency that wider pulses exhibit longer lags (Norris et al. 1996;
Norris, Scargle \& Bonnell 2001; Norris et al. 2005).

\subsubsection{Radius of the radiation surface}

\begin{figure}
\centerline{\hbox{\psfig{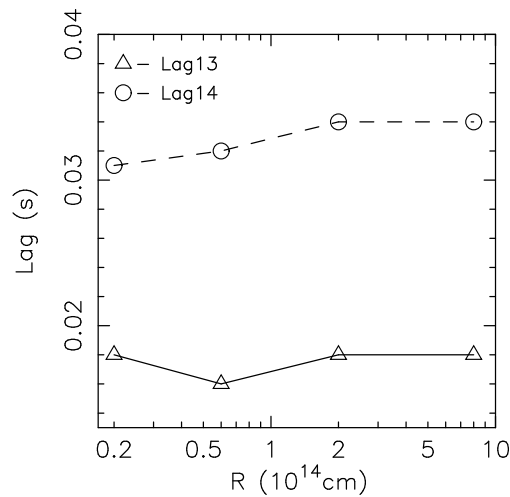}}}
\caption{Symmetric Gaussian intrinsic pulse: lag dependence on
$R$, the radius of radiation surface.  $\Gamma$= 100,
$E_p^{'}$= 1.75 keV, $\alpha$= -0.8, $\beta$= -2.4, $t_d^{'}$= 40 s.}
\label{lag_R}
\end{figure}

The lag appears to be independent on $R$, the radius of the radiation surface
(see Figure \ref{lag_R}). We know $R$ determines the angular spreading time scale,
$T_{ang} \approx R/(2 \Gamma^2 c)$. This result suggests that though in our
model the angular spreading effect is a necessity in causing the lags, the peak
lag in the pulses is not correlated with the angular spreading time scales.

\begin{figure}
\centerline{\hbox{\psfig{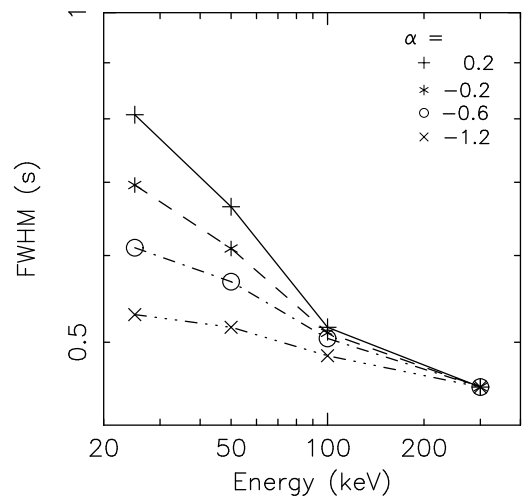}}}
\caption{Symmetric Gaussian intrinsic pulse: pulse width as a
function of the photon energy for different $\alpha$, the low
energy spectral index of the adopted spectrum. Other model
parameters are the same as in Figure \ref{lag_alpha}.}
\label{width_alpha}
\end{figure}

In addition, we find that the decrease of pulse width with the photon energy is
dependent upon the low-energy spectral index of the radiation spectrum, as shown
in Figure \ref{width_alpha}.

%%%%%%%%%%%%%%%%%%%%%%%%%%%%%%%%%%%%%%%%%%%%%%%%%%%%%%%%%%%%%%%
\section{CONCLUSIONS AND DISCUSSIONS}

By assuming an intrinsic ``Band''-shape spectrum and an exponential or Gaussian
emission profile, we show that merely the curvature effect produces detectable soft
lags in the GRB pulses. Therefore the soft time lags can be a signature of the
relativistic motion occurring in GRBs.

The observed Channel 1/3 lags are typically distributed among $10^{-2} - 10^{-1}$ s
(Norris et al. 2000). For typical physical parameters, i.e., $\Gamma \approx$ 100,
$t_d' \approx$ 40 s, $\alpha \approx$ -1, as shown in Figure 4 - 8, the lags produced
by the relativistic curvature effect are slightly above $10^{-2}$ s, marginally close
to those observed, after considering the cosmological time dilation effect
if a GRB redshift of 2 is assumed. To account for those observed larger lags
($\sim$ 0.1 s), it requires extreme physical parameter values, e.g.,
$\Gamma <$ 50 or $\alpha >$ -0.5.

We did not find any peak lag for a rectangular intrinsic
emission profile, from which a straight-forward conclusion regarding the
radiation process of the pulse in GRBs can be obtained ---
the radiation intensity must have a decaying phase in order to produce the observed
peak lags. The intrinsic decaying phase may be
due to the variations associated with hydrodynamic processes, such as the decaying
of emission caused by density or magnetic field attenuation as the shock moves
through the shell.

We have investigated the possible dependences of the pulse peak lag upon other physical
parameters of the kinematic model. We found: the lag is proportional to the inverse of the
Lorentz factor; the lag is proportional to the duration of the intrinsic radiation $t_d'$;
the lag is weakly dependent on $R$; the lag is larger when larger amount of energy is
concentrated at $E_p$ (larger $\alpha$ or smaller $\beta$).

The pulse width decreases with energy ($W \propto E^{-0.1\sim -0.2}$), but not as
fast as the observed ($W \propto E^{-0.4}$ ), though we found a faster decrease with
a larger low-energy spectral index $\alpha$. There must be other energy-dependent
narrowing mechanisms underlying. Similar to this conclusion, Dermer (2004) pointed
out that other processes including adiabatic and radiative cooling, a non-uniform jet,
or the external shock process, rather than the curvature effect, should be needed to
explain the relationship between the measured peak photon energy $E_p$ and the measured
$\nu F_{\nu}$ flux at $E_p$ in the decaying phase of a GRB pulse. They found the
curvature relationship does not agree with the observation (Borgonovo \& Ryde 2001).

For simplicity we have assumed a spherically symmetric radiation surface in this paper,
though there is some observational evidence indicating that the GRB outflow may be
collimated. Derived from the afterglow observations, the jet coming from
the GRB central source generally has a half opening angle $\theta_j > \Gamma^{-1}$
(Frail et al. 2001). Assuming a jet geometry with the jet opening angle of
$1^{\circ}$ or $4^{\circ}$ and the same parameter values ($\Gamma$, $R$ and $t_d$)
used in the spherical geometry, we calculated the observed pulse shapes and the lags;
they show no difference from those of the spherical geometry.
The reason for this is as follows. Even in the case of the isotropic radiation surface,
the contribution of the flux from the outer side region where the
observing angle $\theta$ is larger than $\Gamma^{-1}$ is relatively very small, because
the local flux contribution from the radiation surface (i.e., the first-part integrand of
Equation 5 in Section 2) will decrease drastically with
$\theta$ as $\propto(1+\Gamma^2\theta^2)^{-5}$ when $\theta$ is small.

%%%%%%%%%%%%%%%%%%%%%%%%%%%%%%%%%%%%%%%%%%%%%%%%%%%%%%%%%%%%
\section*{Acknowledgments}
We thank the referee for their careful reading of the manuscript and for many valuable
suggestions. This work was supported in part by the Special Funds for Major State Basic
Science Research Projects of the Ministry of Science and Technology, and by the National
Natural Science Foundation of China through grant 10473010.

\newpage

\end{document}